\documentclass[11pt]{article}

\usepackage[final]{acl}

\usepackage{times}
\usepackage{latexsym}

\usepackage[T1]{fontenc}

\usepackage[utf8]{inputenc}

\usepackage{microtype}

\usepackage{inconsolata}

\usepackage{graphicx}

\usepackage{pgfplots}
\usepackage{subcaption}
\pgfplotsset{compat=1.18}
\usepackage{enumitem}

\usepackage{hyperref}       

\usepackage{url}            
\usepackage{booktabs}       
\usepackage{algorithm}
\usepackage{multirow}
\usepackage{algpseudocode}
\usepackage{algorithmicx}
\usepackage{float}
\usepackage{amsmath}
\usepackage{amsfonts}       
\usepackage{nicefrac}       
\usepackage{xcolor}         
\usepackage{listings}
\usepackage{tcolorbox}
\tcbuselibrary{listings, skins}
\usepackage[capitalize]{cleveref}       \crefname{figure}{Figure}{Figures}
\Crefname{figure}{Figure}{Figures}

\newcommand{\prover}{\textit{HybridProver}}

\definecolor{pinkkeyword}{HTML}{FF66B3} 
\definecolor{skyblue}{HTML}{66b2ff} 

\lstdefinelanguage{Isabelle}{
    morekeywords=[1]{locale, fixes, and, record, assumes, datatype, begin, end, 
    interpretation, lemma, theorem, definition, where, using, by, else, if, then, let, in, SOME,
    proof, next, qed, done, apply, apply, case, show}, 
    sensitive=false,
    morecomment=[l]{\#},
    keywordstyle=[1]\ttfamily\color{pinkkeyword}, 
    basicstyle=\ttfamily\color{black},
    mathescape=true,
    escapeinside={\%*}{*)},
    columns=fullflexible,
    numbers=none,
    xleftmargin=0em,
    keepspaces=true,
    breaklines=true,
    breakautoindent=false,
    breakindent=0pt,
    literate=*
            {_}{\textunderscore}1, 
    escapeinside={||},
    literate=*
        {_}{\textunderscore}1
        |{{\ttfamily\color{skyblue}}}1 %
}

\newtcblisting[auto counter, list inside=mycode]{mylisting}[2][]{ 
    boxsep=0.5mm,
    left=1mm,
    top=-1.5mm,
    bottom=-2mm,
    colback=gray!10,
    colframe=black!60,
    arc=3pt,
    boxrule=1.5pt,
    enhanced,
    drop fuzzy shadow,
    title={\textbf{Listing \thetcbcounter:} #2}, 
    label={listing:\thetcbcounter}, 
    list entry={\textbf{Listing \thetcbcounter} #2}, 
    listing only,
    listing options={
        basicstyle=\ttfamily\color{black},
        mathescape=true,
        escapeinside={\%*}{*)},
        columns=fullflexible,
        numbers=none,
        xleftmargin=0em,
        keywordstyle=\color{pinkkeyword},
        keepspaces=true,
        breaklines=true,
        breakautoindent=false,
        breakindent=0pt,
        literate=*
            {_}{\textunderscore}1 
    },
    #1 
}

\title{HybridProver: Augmenting Theorem Proving with LLM-Driven Proof Synthesis and Refinement}

\author{
  \textbf{Jilin Hu\textsuperscript{1,2}},
  \textbf{Jianyu Zhang\textsuperscript{2}},
  \textbf{Yongwang Zhao\textsuperscript{2,3,*}},
  \textbf{Talia Ringer\textsuperscript{4}}
  \\
  \textsuperscript{1}Shanghai Artificial Intelligence Laboratory,
  Shanghai, China
  \\
  \textsuperscript{2}School of Cyber Science and Technology,
  College of Computer Science and Technology,\\
  Zhejiang University, Hangzhou, China
  \\
  \textsuperscript{3}State Key Laboratory of Blockchain and Data Security,
  Zhejiang University, Hangzhou, China
  \\
  \textsuperscript{4}Siebel School of Computing and Data Science,
  University of Illinois Urbana-Champaign,\\
  Champaign, United States
  \\
  {\small
  \textsuperscript{*}\textbf{Corresponding author:}
  \texttt{zhaoyw@zju.edu.cn}}
}

\begin{document}
\maketitle
\begin{abstract}
Formal methods play a crucial role in ensuring the reliability of critical systems through rigorous mathematical verification. However, their adoption remains limited due to the labor-intensive nature of manual proof construction. Recent advances in large language models (LLMs) have opened new opportunities for automated theorem proving. Two main paradigms have emerged: stepwise tactic-based generation and whole-proof synthesis. While both approaches have complementary strengths, existing work largely treats them in isolation.
In this work, we propose \textit{HybridProver}, a unified framework that integrates whole-proof synthesis and tactic-based generation through \emph{proof sketches} as an intermediate representation. This design enables the reuse of partially correct proof structures while effectively combining high-level planning with fine-grained reasoning.
We implement \textit{HybridProver} in Isabelle/HOL and post-train two 7B-scale LLMs on our optimized Isabelle datasets. Experiments on the miniF2F Isabelle benchmark achieved a 73.8\% success rate and improved upon the previous state of the art (61.9\%), demonstrating that lightweight models, when combined with our approach, can effectively generate Isabelle/HOL proofs without relying on very large LLMs. Ablation studies further analyze the impact of dataset quality, training configurations, and sampling strategies on proof generation. All code and datasets are released.

\end{abstract}

\section{Introduction}
\label{sec:intro}

The field of formal methods plays a pivotal role in ensuring the reliability of critical systems across mathematics, computer science, and engineering~\cite{formalmethods}. 
By constructing machine-checkable proofs over formal specifications, interactive theorem provers (ITPs) provide strong guarantees of correctness. However, the process of developing formal proofs remains labor-intensive and requires substantial expertise, limiting the broader adoption of formal verification.
To address this challenge, automated theorem proving has emerged as a paradigm to streamline formal verification~\cite{barrett2011cvc4,de2008z3,1218615.1218620,weidenbach1996spass}, but its large proof-search space and limited adaptability to diverse proof scenarios have constrained its practicality. 

Recent advances in large language models (LLMs) have opened new frontiers for automated theorem proving. 
Generating formal proofs can be viewed as a structured form of code generation, a setting in which LLMs have demonstrated strong capabilities~\cite{chen2021codex,jiang2024surveylargelanguagemodels}.
Moreover, unlike natural language outputs, formal proofs can be mechanically verified by ITPs, e.g., Isabelle/HOL \cite{nipkow2002isabelle}, Rocq \cite{sozeau2025correct}, and Lean \cite{moura2021lean}, providing a principled way to mitigate hallucination through external verification.

Existing LLM-based approaches to automated theorem proving typically fall into two categories:
(1) whole-proof synthesis that produces complete formal proof attempts in a single pass~\cite{xin2024deepseek,NEURIPS2025_c95c9a9e,wang2024theoremllama}, and (2) tactic-based generation that constructs proofs via stepwise reasoning \cite{zhao2024subgoalxl,jiang2023draft,10.1145/3798275}. 
Whole-proof synthesis excels at high-level planning but is error-prone at the level of detail, while tactic-based generation does the opposite. This suggests an opportunity to reuse the global structure of failed whole-proof attempts while repairing their local errors with tactic-level refinement.

In this work, we present \prover, an LLM-driven framework that integrates these two paradigms through an intermediate representation in the form of proof sketches. 
Given a theorem, \prover\ first generates candidate proofs using a whole-proof synthesis model. When these candidates are incorrect, \prover\ abstracts them into proof sketches that retain the high-level structure while omitting low-level details. These sketches are then verified and refined using a tactic-based generation model, together with existing symbolic tools such as hammers~\cite{blanchette2016hammering}, to produce a final whole proof.

This design highlights a general principle for LLM-based formal reasoning: separating high-level planning from low-level refinement and bridging them through verifiable proof sketches. Such sketches can still be checked for structural validity, enabling tactic-based generation to proceed under a well-formed proof context while reusing partially correct whole-proof attempts. As a result, \prover\ can flexibly compose different models and tools, combining neural generation with symbolic reasoning in a unified framework. More broadly, \prover\ suggests a general pattern for verifier-guided generation: reuse rejected outputs as structured intermediates and repair only the invalid parts. This idea may also apply to structured reasoning and code generation with verifiers.


While the ideas for \prover\ can be implemented for many ITPs, we focus on Isabelle/HOL \cite{nipkow2002isabelle},
using Sledgehammer~\cite{PAAR-2010:Three_Years_Experience_with} as the hammer. 
Compared to prior LLM-based theorem-proving systems, which predominantly focus on Lean, Isabelle presents a complementary and relatively underexplored setting.
It poses additional challenges for LLM-based proof generation due to its diverse proof styles and less standardized proof patterns, while also offering unique opportunities for constructing high-quality training data (see Section~\ref{gen_inst}).

We evaluate \prover\ on the miniF2F benchmark~\cite{zheng2022minif2f}. Our approach improves proof success rates over prior Isabelle-based systems. We further show that combining whole-proof synthesis with tactic-based generation yields consistent performance gains, and that integrating symbolic tools such as Sledgehammer yields additional improvements.
Beyond overall performance, our ablation studies highlight the importance of dataset quality, fine-tuning hyperparameters, and sampling diversity in LLM-based proof generation.
To facilitate future research, we release our datasets and the code of \prover.\footnote{\url{https://github.com/JilinHu/HybridProver}}

\paragraph{Contributions:}
\begin{itemize}
    \item We propose \prover, a framework for LLM-based automated theorem proving that unifies whole-proof synthesis and tactic-based generation. By using proof sketches as intermediate representations, \prover\ connects high-level proof planning with low-level tactic refinement. It also enables imperfect whole-proof candidates to be reused and refined rather than discarded.
    
    
   \item We implement \prover\ in Isabelle/HOL and construct optimized training datasets. We further fine-tune two 7B-scale LLMs for both whole-proof synthesis and tactic-based generation, and enhance them with reinforcement learning guided by verification signals.
    
   \item \prover\ achieves a 73.8\% success rate on miniF2F, improving over prior results for Isabelle-based systems (61.9\%). We demonstrate that lightweight models can be effectively adapted to proof generation through high-level planning and low-level tactic refinement. Extensive ablations highlight the impact of data quality, fine-tuning hyperparameters, and sampling diversity.

\end{itemize}

\section{Background: Why Isabelle?}
\label{gen_inst}

We implement \prover\ in Isabelle/HOL, an interactive theorem prover (ITP) widely used for formal verification and mathematical reasoning~\cite{nipkow2002isabelle}. 
ITPs such as Isabelle, Rocq~\cite{sozeau2025correct}, and Lean~\cite{moura2021lean} provide strong correctness guarantees by checking proofs with small, trusted kernels~\cite{nipkow2002isabelle,pierce2010softwarefoundation}. 
However, constructing formal proofs in these systems remains labor-intensive and requires significant expertise~\cite{klein2009sel4,gu2016certikos,leroy2016compcert}. 
Our choice of Isabelle as a target ITP addresses its unique challenges while also taking advantage of its unique benefits.

\begin{figure*}[t]
  \centering
  \includegraphics[width=\textwidth]{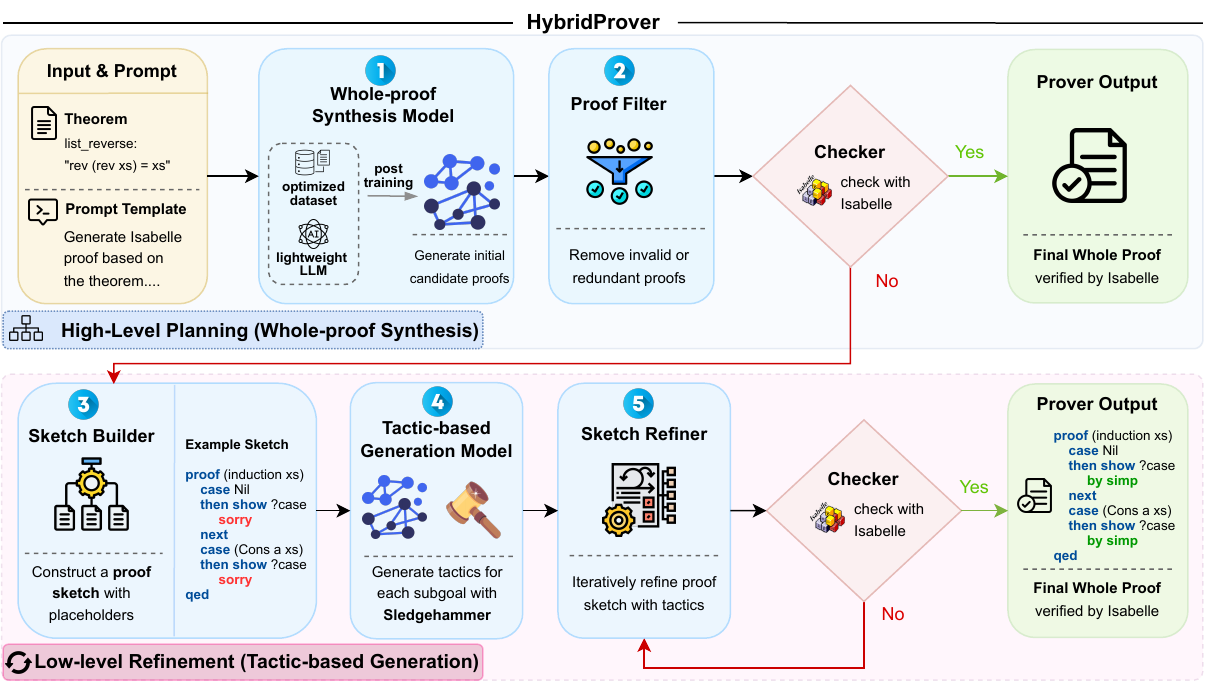}
  \caption{HybridProver Framework Overview}
  \label{fig:overview}
\end{figure*}

\paragraph{Challenges of targeting Isabelle: broad use cases.}


Compared to prior LLM-based theorem proving systems, which predominantly focus on Lean, Isabelle remains relatively underexplored. 
This difference is partly due to the nature of existing benchmarks such as miniF2F, which are mathematical and better aligned with Lean’s mathlib~\cite{moura2021lean}. 
In contrast, Isabelle is more widely used in system verification, where proofs vary significantly across domains and specifications. 
While Isabelle provides a large and diverse corpus of formal proofs (over 200K theorems, compared to Lean's 140K and Rocq's 71K~\cite{drori2025diverseinferenceverificationadvanced,reichel2023proof}). This diversity makes it challenging for LLMs to learn consistent proof patterns across domains.

\paragraph{Advantages of targeting Isabelle: two proof styles.}

A distinctive feature of Isabelle is the coexistence of two proof styles: apply-style proofs~\cite{IsabelleTutorial} and Isar proofs~\cite{IsarRef}. 
Apply-style proofs construct derivations through stepwise tactic application, where a tactic is a primitive proof step that refines the current proof state to an updated state. In contrast, Isar proofs provide structured, human-readable proof scripts that resemble mathematical reasoning. 

\begin{mylisting}[listing options={language=Isabelle,basicstyle=\ttfamily\footnotesize}]{Isar and apply-style proof example}
theorem list_reverse: "rev (rev xs) = xs"
|\ttfamily\color{skyblue}Isar proof:|             |\ttfamily\color{gray}\textbar||\ttfamily\color{skyblue}apply-style proof:|
proof(induction xs)     |\ttfamily\color{gray}\textbar|apply(induction xs)
case Nil                |\ttfamily\color{gray}\textbar| apply auto
then show ?case by simp |\ttfamily\color{gray}\textbar| done
next case (Cons a xs)   |\ttfamily\color{gray}\textbar|
then show ?case by simp |\ttfamily\color{gray}\textbar|
qed                     |\ttfamily\color{gray}\textbar|


\end{mylisting}

Listing~\ref{listing:1} illustrates these two styles on the same theorem, which states that reversing a list twice does not change the list.
Apply-style proofs are concise but opaque, requiring inspection of intermediate proof states, whereas Isar proofs explicitly expose the high-level structure. 

Importantly, these two styles are complementary: Isar can serve as a high-level proof sketch, while apply-style tactics can discharge low-level subgoals. This observation motivates our design, where we treat Isar-style proofs as structured sketches and use tactic-based generation to refine them. 
By leveraging both styles, we construct a unified framework that bridges high-level planning and low-level reasoning, and enables more effective proof generation.

\section{HybridProver}

\subsection{Framework Overview}
\label{sec:framework}
Our framework is depicted in \cref{fig:overview}. The workflow can be divided into the following steps:

\textbf{Step 1: Prompt Construction.}
\prover\ takes as input a formal theorem statement in Isabelle syntax.
The statement is concatenated with a structured prompt template (Listing~\ref{listing:prompt}) that specifies the task and output format, and is then fed into a post-trained \textit{whole-proof synthesis model}.
Trained on optimized Isabelle proof data, the model generates candidate proofs by capturing common proof patterns and basic reasoning structures.


\textbf{Step 2: Whole-Proof Synthesis and Filtering.}
The whole-proof synthesis model generates $n$ candidate proofs (we set $n=128$ following prior work) in parallel to explore diverse proof paths.
A lightweight \textit{proof filter} is then applied to remove invalid or redundant outputs, reducing the cost of subsequent verification.
The filter enforces basic syntactic constraints on Isabelle proofs. 
Outputs that do not conform to the required JSON format or do not start with valid Isabelle keywords (e.g., \textit{proof}, \textit{apply}, or \textit{by}) are discarded.
The remaining candidates are passed to the Isabelle checker for formal verification, noting that syntactic validity does not guarantee correctness.

With concurrent verification, multiple candidates can be checked simultaneously.
If any candidate passes verification, the system returns the corresponding proof and terminates early.
Otherwise, the valid candidates are forwarded to the next stage for further processing.

\textbf{Step 3: Proof Sketch Construction.}
If none of the $n$ candidate proofs passes the checker, the system invokes a \textit{sketch builder} to transform syntactically valid candidates into proof sketches, rather than discarding them.
The key idea is to lift detailed proof attempts to a higher-level representation that preserves global structure while abstracting away low-level tactic details.

Unlike prior work~\cite{jiang2023draft}, we construct proof sketches directly from LLM-generated candidate proofs.
The sketch builder validates each subgoal proof independently: when checking one subgoal, it replaces the proofs of the remaining subgoals with \textit{sorry}, allowing Isabelle to verify the current proof step under assumed completion of the others.
Correct subgoal proofs are preserved, while failed ones are replaced with \textit{sorry}.
The resulting sketch is therefore a partially verified intermediate representation that reuses correct local proof steps and leaves only unresolved subgoals for refinement.
The resulting sketches are then validated by the Isabelle checker to remove structurally invalid cases, such as incorrect proof methods or missing lemmas.
Since low-level details are abstracted away, most sketches pass this stage.
With concurrent verification, this process remains efficient.

\textbf{Step 4: Subgoal Refinement and Tactic Generation.}
Given a proof sketch, we refine each subgoal marked with \textit{sorry} into a verifiable proof step using a \textit{tactic-based generation model}.
This model combines a post-trained LLM for tactic generation with the Isabelle tool Sledgehammer.

For each unresolved subgoal, the LLM generates $n$ candidate tactics in parallel, while Sledgehammer independently attempts to discharge the same subgoal.
The resulting candidates are validated by the Isabelle checker.
These two components are complementary, as each can solve subgoals that the other fails to prove.
In contrast to whole-proof synthesis, which focuses on high-level structure and premise selection, this stage operates on localized subgoals and concrete proof steps.
In our framework, premise selection is incorporated into the training of the whole-proof synthesis model, allowing the refinement stage to leverage richer contextual information for subgoal solving.

\textbf{Step 5: Proof Refinement.}
In the final stage, the \textbf{sketch refiner} replaces the \textit{sorry} placeholders in the proof sketch with the corresponding proof steps generated in Step 4, yielding a complete Isabelle proof.
The refined proof is then submitted to the Isabelle checker for final verification.
If verification fails, the system performs backtracking by substituting alternative tactics for unresolved subgoals and re-verifying the reconstructed proof.
This process repeats until a valid proof is found or all candidates are exhausted.
If successful, the system returns the verified proof; otherwise, it reports failure.
This iterative refinement process bridges the gap between abstract proof sketches and complete proof code.

\begin{algorithm}[t]
\caption{HybridProver workflow}
\label{alg:atp}
\footnotesize
\setlength{\textfloatsep}{4pt}
\begin{algorithmic}[1]
\Require Theorem $L$, prompt template $T$
\Ensure Valid Isabelle proof or failure

\State $\mathcal{P} \gets \mathrm{LLM}_{\mathit{whole}}(L,T)$ \hfill \textit{// Step 1: whole-proof synthesis}
\State $\mathcal{P}_{\mathit{syntax}} \gets \mathit{proof\_filter}(\mathcal{P})$ \hfill \textit{// Step 2: filtering}
\State $\mathcal{P}_{\mathit{valid}} \gets \emptyset$
\State \textbf{parallel for} $p \in \mathcal{P}_{\mathit{syntax}}$ \textbf{do}
\If{$\mathrm{check}(p,L)$}
        \State $\mathcal{P}_{\mathit{valid}} \gets \mathcal{P}_{\mathit{valid}} \cup \{p\}$
    \EndIf \ \textbf{end parallel} 
\If{$\mathcal{P}_{\mathit{valid}} \neq \emptyset$}
    \State \Return any $p \in \mathcal{P}_{\mathit{valid}}$
\EndIf
\State $\mathcal{S} \gets \{\mathrm{replaceSorry}(p)\mid p \in \mathcal{P}\}$ \hfill \textit{// Step 3: sketch construction}
\State $\mathcal{S}_{\mathit{valid}} \gets \{s \in \mathcal{S}\mid \mathrm{check}(s,L)\}$

\State \textbf{parallel for} $s \in \mathcal{S}_{\mathit{valid}}$ \textbf{do} \hfill \textit{// Steps 4--5: refinement}
    \State $G_1,\dots,G_n \gets \mathrm{parseSubgoals}(s)$
    \State $\mathcal{T} \gets \bigcup_{i=1}^{n}\bigl(\mathrm{LLM}_{\mathit{step}}(G_i)\cup \mathrm{Sledgehammer}(G_i)\bigr)$
    \State $p_{\mathit{refined}} \gets \mathrm{substituteSorry}(s,\mathcal{T})$
    \If{$\mathrm{IsabelleVerify}(p_{\mathit{refined}},L)$}
        \State \Return $p_{\mathit{refined}}$
    \EndIf \ \textbf{end parallel} 

\State \Return failure
\end{algorithmic}
\end{algorithm}

The overall workflow of \prover\ is summarized in \cref{alg:atp}.
Given a theorem $L$ and a prompt template $T$, the whole-proof synthesis model generates candidate proofs, which are filtered and verified in parallel.
If a valid proof is found, the system terminates early.
Otherwise, syntactically valid candidates are converted into proof sketches and validated.
For each valid sketch, subgoals are extracted and solved using a tactic-based generation model together with Sledgehammer.
The resulting tactics are used to refine the sketch into a complete proof, which is then verified by Isabelle.
This process iterates until a valid proof is found or all candidates are exhausted. An Isabelle running example is shown in \cref{append:running example}.

\subsection{Dataset optimization}
\label{sec:datasetoptimization}

We construct our Isabelle dataset from the Standard Library (STD) and the Archive of Formal Proofs (AFP) using PISA~\cite{jiang2021lisa}.
To reduce noise in raw formalization data, we apply a cleaning pipeline that extracts theorem–proof pairs, removes formatting artifacts (e.g., redundant whitespace), eliminates comments, and normalizes symbols.
Each proof is then converted into a structured format, \textit{\{Isabelle\_proof: proof code\}}, to standardize the target output.
We further partition the dataset according to Isabelle’s two proof styles: Isar and apply-style proofs.
This distinction aligns with our two-stage framework: Isar proofs capture high-level structure, while apply-style proofs focus on tactic-level reasoning.

To improve training efficiency and data quality, we filter examples based on proof length.
Apply-style proofs in raw data are typically short (1–5 steps), so we retain only concise instances to encourage precise tactic generation.
Isar proofs are longer (typically 11–30 steps), and we remove overly short or excessively long examples (outside 5–50 steps) to balance structural richness and learnability.
The resulting dataset provides a cleaner and more structured training signal for both global proof generation and local tactic refinement.

\subsection{Model Training}

We first fine-tune two models corresponding to the two stages of \prover.
The whole-proof synthesis model is trained on the Isar dataset to capture high-level proof structure, while the tactic-based generation model is trained on the apply-style dataset to generate fine-grained proof steps.
Both models are obtained via supervised fine-tuning of pretrained models.
To improve training efficiency and reduce memory consumption, we adopt LoRA~\cite{hu2022lora}, which freezes the base model and introduces low-rank trainable adapters.
This allows us to scale to larger models while maintaining performance comparable to full fine-tuning under limited computational resources.

After supervised fine-tuning, we further optimize the model using reinforcement learning. We adopt Group Relative Policy Optimization (GRPO)~\cite{shao2024deepseekmathpushinglimitsmathematical}, where Isabelle serves as the reward provider. The training data consist of a randomly sampled 1\% subset of the optimized dataset. A key advantage of theorem proving is that generated proofs are formally verifiable, making Isabelle an accurate and reliable reward signal. Specifically, proofs that pass verification receive the highest reward, partially correct but invalid proofs receive a lower reward, and syntactically invalid or rejected outputs receive no reward. We additionally introduce penalties to discourage degenerate behaviors, such as generating placeholders (e.g., \texttt{sorry}) or trivial outputs. The final reward combines verification feedback with these structural penalties. This design encourages the model to generate complete, verifiable proofs while avoiding shortcuts that bypass actual reasoning.



\section{Experiments}
\label{sec:experiment}







\subsection{Setup}
\label{sec:setup}

\paragraph{Dataset.}

We construct the training dataset from the Isabelle Standard Library (STD) and the Archive of Formal Proofs (AFP) using PISA \cite{jiang2021lisa}, following the preprocessing described in \cref{sec:datasetoptimization}. 
AFP is split into 95\%/1\%/4\% for training/validation/test following prior work~\cite{jiang2022thor,jiang2023draft}, while STD is used entirely for training due to its dependency on AFP. 
The dataset statistics are summarized in \cref{tab:dataset_source_configuration}.
For evaluation, we use the miniF2F benchmark \cite{zheng2022minif2f}, a widely used dataset consisting of 488 mathematical problems. 
Following prior work, we report results on the 244 test problems. miniF2F is used exclusively for evaluation in our pipeline and is not included in either SFT or GRPO data; all post-training examples are drawn from AFP and STD. The evaluation inputs provide theorem statements rather than target proofs, which substantially limits direct proof-target leakage.

\begin{table}[t]
\centering
\caption{Dataset Sources and Configurations}
\label{tab:dataset_source_configuration}
\small
\setlength{\tabcolsep}{4pt}
\renewcommand{\arraystretch}{1.1}
\begin{tabular}{cccc}
\hline
\textbf{Data Source} & \textbf{Purpose} & \textbf{Proof Style} & \textbf{Samples} \\
\hline
\multirow{4}{*}{AFP} 
& \multirow{2}{*}{Train (95\%)} & apply-style & 116,308 \\
&  & Isar-style & 35,803 \\
\cline{2-4}
& Test (4\%) & -- & 6,400 \\
& Validation (1\%) & -- & 1,600 \\
\hline
\multirow{2}{*}{STD}
& \multirow{2}{*}{Train (100\%)} & apply-style & 36,038 \\
&  & Isar-style & 9,018 \\
\hline
\end{tabular}
\end{table}


\paragraph{Experimental setup.}
In initial small-model experiments, we generated proofs using more than 15 different open-source models. In the end, we chose Qwen-Coder-7B as our base model for further training because it had the highest proof success rate. For LLM fine-tuning and reinforcement learning (RL), we use 8 NVIDIA A40 GPUs. Fine-tuning our models with LoRA on the whole optimized Isabelle dataset and 3 epochs takes around 200 GPU hours, while RL with GRPO takes around 100 GPU hours. The cost of ablation studies for other LLMs and configurations is not included. Detailed training configurations can be found in \cref{append:sft and rl}.
For evaluation, we use a 128-core Intel CPU with 256 GB of RAM while setting the concurrent processes to 64. The full evaluation required nearly 10,000 CPU hours. At inference time, computation is divided among whole-proof sampling and checking, independent subproof and sketch validation, tactic generation, Sledgehammer, and repeated verification during backtracking.

\subsection{Results and Analysis}
\label{sec:results}

We evaluate \prover\ on the Isabelle miniF2F test set, using Sledgehammer as the baseline.
Following prior work, we use a sampling size of 128 when generating candidate proofs.
The results are shown in \cref{tab:minif_success}.
Sledgehammer with heuristics achieves a 20.9\% success rate without LLM assistance.
Recent LLM-based approaches significantly improve over this baseline, reaching 30\%--50\% success rates.
With the integration of whole-proof synthesis and tactic-based generation, \prover\ achieves a success rate of 73.8\%, outperforming prior Isabelle-based systems and improving over the previous best result of 61.9\% from ProofAug~\cite{10.5555/3780338.3781921} by 11.9\%.

\begin{table}[!htbp]
\centering
\caption{Proof success rates on the Isabelle miniF2F}
\label{tab:minif_success}
\small
\setlength{\tabcolsep}{4pt}
\begin{tabular}{lcc}
\toprule
\textbf{Method (using Isabelle)} & \textbf{Base Model} & \textbf{Test (\%)} \\
\midrule
Sledgehammer & -- & 10.4 \\
Sledgehammer + heuristics & -- & 20.9 \\
\midrule
Thor & -- & 29.9 \\
Thor + expert iteration & -- & 35.2 \\
\midrule
DSP & Codex & 39.3 \\
POETRY & ProofGPT(1.3B) & 42.2 \\
Subgoal-Prover & GPT-3.5-Turbo & 45.5 \\
LEGO-Prover & GPT-3.5-Turbo & 50.0 \\
Lyra & GPT-4 & 51.2 \\
SubgoalXL & Llama3-8B & 56.1 \\
ProofAug & Deepseek-math-7B & 61.9 \\
\midrule
\textbf{HybridProver} & \textbf{Qwen-Coder-7B} & \textbf{73.8} \\
\bottomrule
\end{tabular}
\end{table}

To analyze the contribution of each component, we evaluate several variants of \prover\ (\cref{tab:model training}).
The whole-proof synthesis model alone achieves 45.9\%, while the tactic-based generation model achieves 39.3\%.
A simple combination of the two models (taking the union of their results) improves the success rate to 49.6\%.
In miniF2F, some theorems are simple and require only a small number of proof steps.
In such cases, the whole-proof synthesis model tends to generate verbose and often incorrect Isar proofs, while the tactic-based generation model is more effective at producing concise solutions.
This explains the complementary behavior of the two models.
Integrating Sledgehammer with heuristics further improves performance by approximately 2\%, suggesting that the tactic-based LLM already covers most subgoals that can be solved by symbolic tools.

\begin{table}[!htbp]
\centering
\caption{Ablation results on Isabelle miniF2F}
\label{tab:model training}
\footnotesize
\setlength{\tabcolsep}{2.5pt}
\renewcommand{\arraystretch}{1.03}
\begin{tabular}{p{0.70\columnwidth}r}
\toprule
\textbf{Variant} & \textbf{Test (\%)} \\
\midrule
Whole-proof synthesis model & 45.9 \\
Tactic-based generation model & 39.3 \\
\midrule
Whole-proof + tactic generation & 49.6 \\
Tactic generation + sketch refinement & 71.3 \\
\midrule
\textbf{HybridProver} & \textbf{73.8} \\
\bottomrule
\end{tabular}
\end{table}

Combining tactic-based generation with proof sketch refinement significantly boosts performance to 71.3\%, demonstrating the effectiveness of reusing partially correct proofs.
Interestingly, most sketches constructed from the whole-proof synthesis model are structurally correct: when evaluated without detailed tactics, sketches achieve up to 90\% success rate.
This indicates that high-level proof structure is often captured correctly even when local reasoning fails.
Finally, \prover\ achieves a 73.8\% success rate, showing that combining whole-proof synthesis with tactic-based refinement leads to substantial improvements, while symbolic tools provide additional, though smaller, gains.
Besides, lightweight models can effectively generate Isabelle/HOL proofs when combined with our approach without relying on very large LLMs. Qualitatively, remaining failures can arise from malformed or unparseable Isar structure, hallucinated or missing premises, unsuitable local tactics and exhaustion of retained tactic combinations in multi-subgoal sketches.

In our pipeline, SFT adapts the models to the Isabelle output format and the Isar/apply-style proof patterns, while GRPO subsequently optimizes them using Isabelle-derived verification rewards. However, the current ablations do not hold all other factors fixed to isolate the individual gains from SFT and GRPO. We therefore attribute the observed improvements to the post-training pipeline as a whole rather than to either stage separately.

\subsection{Additional Ablation Studies}
\label{sec:ablation}

\paragraph{RQ1: How do data quality and training parameters affect fine-tuning performance?}

We use the original Qwen-Coder-7B model as a baseline. 
Since it is only generically pretrained, it struggles to follow our task-specific instructions.
To mitigate this, we prompt the model with a structured format (Listing~\ref{listing:prompt}) and enforce constrained outputs.

We then fine-tune the whole-proof model on variants of our optimized Isabelle dataset under different hyperparameter settings, including learning rate, LoRA rank, and number of training epochs.
Our results show that data quality, learning rate, and training duration have a significant impact on performance.
In particular, training on raw data leads to noisy outputs with redundant tokens and formatting artifacts, which degrades proof generation quality.
After applying the data cleaning procedure described in \cref{sec:datasetoptimization}, the success rate on miniF2F improves by at least 10\% across all training settings, highlighting the importance of dataset quality.

The ablation results for training parameters are shown in \cref{fig:training_sampling_ablation}. For training epochs, we find that 3 epochs yield the best performance, while longer training leads to overfitting.
This is partly due to the mismatch between the training data (covering mathematics, software, and systems) and the evaluation benchmark (miniF2F, which is purely mathematical).
For the learning rate, we experiment with values ranging from $10^{-4}$ to $5\times10^{-7}$ using a warm-up and cosine decay schedule, and find that $10^{-5}$ yields the best performance.
Finally, we evaluate different LoRA ranks (8–256) and observe that the rank has minimal effect on performance, despite changing the number of trainable parameters.
These findings provide practical guidance for fine-tuning LLMs in theorem proving tasks.

\begin{tcolorbox}[
  colback=gray!10,
  colframe=gray!80,
  boxrule=1.5pt,
  arc=4pt,
  left=6pt,
  right=6pt,
  top=6pt,
  bottom=6pt
]
\textbf{RA1:} Data quality and training hyperparameters (e.g., learning rate and epochs) significantly affect performance, while LoRA rank has limited impact.
\end{tcolorbox}

\paragraph{RQ2: Does sampling diversity impact the results?}
We vary sampling parameters during inference, including temperature $T$ and sampling rate. While proof generation requires stable outputs to ensure syntactic correctness, prior work adopts higher temperatures combined with multiple samples to improve success rates (e.g., $T{=}1.0$ in LISA and $T{=}1.2$ in Thor). 
In our setting, we evaluate the trained Qwen-Coder-7B model for whole proofs with $T \in [0.3, 1.2]$ and sampling rates from @32 to @128 (\cref{fig:training_sampling_ablation}). We observe that moderate temperatures ($0.6$–$0.9$) achieve the best performance, while both lower and higher values degrade success rates. Error analysis shows that excessively high $T$ leads to frequent hallucination of non-existent facts, causing verification failures in Isabelle. We also find that increasing the sampling rate improves performance, indicating the benefit of broader exploration during inference.

\begin{figure*}[t]
\centering
\captionsetup[subfigure]{skip=1pt}
\begin{subfigure}[t]{0.35\textwidth}
\centering
\begin{tikzpicture}
\begin{axis}[
    width=0.82\linewidth,
    height=2.3cm,
    scale only axis,
    ylabel={Accuracy (\%)},
    ylabel style={font=\small, yshift=-4pt},
    tick label style={font=\small},
    ymin=20, ymax=46,
    symbolic x coords={$e^{-4}$,$5e^{-5}$,$e^{-5}$,$5e^{-6}$,$e^{-6}$,$5e^{-7}$},
    xtick=data,
    x tick label style={font=\small, rotate=0, anchor=north},
    xticklabel style={yshift=1pt},
    enlarge x limits=0.06,
    grid=major,
]
\addplot+[color=skyblue,mark=*,mark options={fill=skyblue}] coordinates {
    ($e^{-4}$,22)
    ($5e^{-5}$,27.6)
    ($e^{-5}$,34.8)
    ($5e^{-6}$,44.7)
    ($e^{-6}$,43.8)
    ($5e^{-7}$,44.6)
};
\end{axis}
\end{tikzpicture}
\caption{Learning Rate}
\end{subfigure}
\begin{subfigure}[t]{0.35\textwidth}
\centering
\begin{tikzpicture}
\begin{semilogxaxis}[
    width=0.82\linewidth,
    height=2.3cm,
    scale only axis,
    ylabel={Accuracy (\%)},
    ylabel style={font=\small, yshift=-4pt},
    tick label style={font=\small},
    ymin=36, ymax=44,
    xmin=8, xmax=256,
    log basis x=2,
    xtick={8,16,32,64,128,256},
    xticklabels={8,16,32,64,128,256},
    x tick label style={font=\small},
    xticklabel style={yshift=1pt},
    grid=major,
]
\addplot+[color=skyblue,mark=*,mark options={fill=skyblue}] coordinates {
    (8,40.7)
    (16,39.4)
    (32,37.7)
    (64,40.6)
    (128,38.1)
    (256,42.2)
};
\end{semilogxaxis}
\end{tikzpicture}
\caption{LoRA Rank}
\end{subfigure}
\vspace{0.35em}
\hspace*{-0.01\textwidth}
\begin{subfigure}[t]{0.35\textwidth}
\centering
\begin{tikzpicture}
\begin{axis}[
    width=0.82\linewidth,
    height=2.3cm,
    scale only axis,
    ylabel={Accuracy (\%)},
    ylabel style={font=\small, yshift=-4pt},
    tick label style={font=\small},
    ymin=34, ymax=46,
    xtick={1,3,6,10},
    x tick label style={font=\small},
    xticklabel style={yshift=1pt},
    grid=major,
]
\addplot+[color=skyblue,mark=*,mark options={fill=skyblue}] coordinates {
    (1,34.8)
    (3,44.6)
    (6,40.6)
    (10,37.7)
};
\end{axis}
\end{tikzpicture}
\caption{Epoch}
\end{subfigure}
\begin{subfigure}[t]{0.35\textwidth}
\centering
\begin{tikzpicture}
\begin{axis}[
    width=0.82\linewidth,
    height=2.3cm,
    scale only axis,
    ylabel={Accuracy (\%)},
    ylabel style={font=\small, yshift=-4pt},
    tick label style={font=\small},
    ymin=25, ymax=41,
    xtick={0.3,0.6,0.9,1.2},
    x tick label style={font=\small},
    xticklabel style={yshift=1pt},
    grid=major,
    legend style={
        at={(0.5,0.09)},
        anchor=south,
        legend columns=2,
        font=\scriptsize,
        draw=black,
        fill=white,
        inner sep=1.5pt
    },
]
\addplot+[color=pinkkeyword,mark=square*,mark options={fill=pinkkeyword}] coordinates {
    (0.3,26.6)
    (0.6,32.7)
    (0.9,33.1)
    (1.2,30.3)
};
\addlegendentry{@32}


\addplot+[color=skyblue,mark=*,mark options={fill=skyblue}] coordinates {
    (0.3,31.1)
    (0.6,39.3)
    (0.9,39.5)
    (1.2,36.9)
};
\addlegendentry{@128}
\end{axis}
\end{tikzpicture}
\caption{Sampling Temperature}
\end{subfigure}

\caption{Ablation results under different training and sampling configurations.}
\label{fig:training_sampling_ablation}
\end{figure*}

\begin{tcolorbox}[
  colback=gray!10,
  colframe=gray!80,
  boxrule=1.5pt,
  arc=4pt,
  left=6pt,
  right=6pt,
  top=6pt,
  bottom=6pt
]
\textbf{RA2:} Sampling diversity influences results: moderate temperature works best, and higher sampling rates improve success rates.
\end{tcolorbox}

\paragraph{RQ3: Does providing premises, few-shot examples, and labels help with proving?} 
Many Isabelle theorems rely on auxiliary premises. We therefore include premises in the training data and train the model to both select relevant premises and construct proofs accordingly. This reduces \textit{undefined fact} errors by around 50\%, indicating that the model learns to avoid hallucinating nonexistent premises. However, this improvement does not translate into better performance on miniF2F, where most theorems do not require external premises.

For few-shot prompting, we add three representative examples (Listing~\ref{listing:few-shot}) to the prompt, covering simple apply-style proofs, induction proofs, and Isar proofs. We observe no performance gain for either the trained Qwen-Coder-7B or the base model, suggesting that the model already captures the required syntax and proof patterns.
We further augment each theorem with its source file path as a label, aiming to provide coarse-grained contextual information for premise selection. While this reduces the search space in principle (e.g., encouraging the model to retrieve premises from the corresponding theory file), it again yields negligible improvements on miniF2F, likely due to its homogeneous mathematical domain.

\begin{tcolorbox}[
  colback=gray!10,
  colframe=gray!80,
  boxrule=1.5pt,
  arc=4pt,
  left=6pt,
  right=6pt,
  top=6pt,
  bottom=6pt
]
\textbf{RA3:} Additional signals such as premises, few-shot examples, and labels provide limited gains on miniF2F.
\end{tcolorbox}

\section{Related Work}
\label{headings}

The use of LLMs to automate theorem proving was pioneered by Polu and Sutskever \cite{polu2020generative}. 
Subsequent work has focused on improving the reasoning capabilities of LLM-based systems.
Open-source interactive environments like
Leandojo~\cite{yang2023leandojo} for Lean, and PISA \cite{jiang2021lisa} and Isabelle-client \cite{10.1007/978-3-031-16681-5_24} for Isabelle, make it easier to build LLM-based tools for automated theorem proving.
Likewise, open-source datasets and benchmarks like miniF2F \cite{zheng2022minif2f} and Proofnet \cite{azerbayev2023proofnet} provide the basis for evaluation.
Automation often uses premise selection prior to proof generation
to find relevant premises~\cite{yang2023leandojo,jiang2022thor,wang2024lego,mikula2023magnushammer}. For example, LEGO \cite{wang2024lego} builds a growing library and retrieves premises from it, while Thor \cite{jiang2022thor} delegates premise selection to hammers during the proving process. In \prover, we ask both the trained LLM and Sledgehammer in Isabelle to perform premise selection. 

Prior work on LLM-based proof generation also varies by what the
LLM is used for and by its input context.
Some work \cite{yang2023leandojo,zhao2024subgoalxl,polu2020generative,lin2024fvel,NEURIPS2025_c95c9a9e,10.5555/3737916.3740669,achim2025aristotleimolevelautomatedtheorem} uses LLMs to generate tactic-based proofs step by step. Other work \cite{xin2024deepseek,first2023baldur,wang2024theoremllama,ren2025deepseekproverv2advancingformalmathematical,shao2025deepseekmathv2selfverifiablemathematicalreasoning,wang2025kiminaproverpreviewlargeformal,lin2025goedelproverfrontiermodelopensource,chen2025seedproverdeepbroadreasoning,10.1145/3798275} uses LLMs to generate the whole proof at once. 
For example, Polu \cite{polu2020generative} and FVEL \cite{lin2024fvel} ask LLM to generate the next tactic based on the current proof state. Baldur \cite{first2023baldur} makes LLMs generate whole proofs based on the input theorem. \prover\ organically combines these two approaches and gets a better proof quality.
After the proof generation, some works also focus on proof search that chooses appropriate generated responses with different algorithms, such as best-first search~\cite{polu2020generative,han2022proof,yang2023leandojo, xin-etal-2025-bfs}, and Monte Carlo tree search \cite{wang2023dt,lample2022hypertree,xin2024deepseek}.
Despite the improvement, these search algorithms often require additional computational resources that are non-trivial. Besides, a few works try to integrate hammers and sketches to improve the performance \cite{jiang2022thor,jiang2023draft,mikula2023magnushammer,zhao2024subgoalxl,yang2023leandojo,10.5555/3780338.3781921}. For instance, DSP \cite{jiang2023draft} generates proof sketches with subgoals and uses hammers to boost the proof success rate. Recent work~\cite{NEURIPS2025_3b77109a,ding2026fmagentscalingformalmethods} has also explored agent-based approaches to enhance the reasoning capabilities of LLMs.

\prover\ differs from them in three aspects. First, most prior work focuses on Lean, while we study another important ITP called Isabelle. Second, we not only integrate hammers but also post-train LLMs based on our optimized dataset to improve the performance. Third, we build the proof sketch mechanically from the whole proof, which bridges the gap between whole-proof synthesis and tactic-based generation, while other work, such as DSP, builds sketches from informal proofs by prompting another LLM.

\section{Conclusion and Future Work}
\label{sec:discussion}
We introduced \prover, an LLM-driven framework for automated theorem proving that unifies whole-proof synthesis and tactic-based generation through proof sketches as an intermediate representation. This design enables effective interaction between global planning and local reasoning, further enhanced by automated tools such as hammers.
In addition, we construct optimized Isabelle datasets and train LLMs via both supervised fine-tuning and reinforcement learning guided by formal verification signals. Our implementation of \prover\ in Isabelle achieves a proof success rate of 73.8\% on miniF2F under our evaluation setting.

In future work, we plan to integrate more advanced proof search techniques into the tactic-based generation process to improve efficiency. Beyond miniF2F, we plan to evaluate \prover\ on broader Isabelle/HOL domains using carefully defined AFP-based evaluation sets. We also plan to develop a more user-friendly integrated system to facilitate practical adoption.


\section*{Limitations}

One limitation of \prover\ lies in its relatively simple proof search strategy during the proof sketch refinement stage. As the number of subgoals increases, the refinement and evaluation process becomes increasingly inefficient, leading to substantial computational overhead.
Moreover, \prover\ relies heavily on the underlying theorem prover (e.g., Isabelle) as a verification oracle. While this provides accurate and reliable feedback, it also introduces a bottleneck in terms of latency and scalability, especially when evaluating a large number of candidate proofs.
Another limitation is that the model is trained on existing proof corpora, which may bias it toward familiar proof patterns. As a result, its ability to generalize to novel or significantly different proof structures remains limited. In addition, our empirical conclusions are limited to miniF2F and do not establish generalization across broader Isabelle/HOL domains.

\bibliography{custom}

\appendix

\section{Technical Appendices and Supplementary Material}
\label{appendix}
\subsection{Hammers}
We integrate hammers in our framework to augment theorem proving.
Blanchette et al. \cite{blanchette2016hammering} define hammers as methods that ``automate reasoning over large libraries developed with formal proof assistants.'' Most ITPs have their own hammers, e.g., Coqhammer in Rocq and Sledgehammer in Isabelle. Sledgehammer in Isabelle is the most powerful one \cite{PAAR-2010:Three_Years_Experience_with}. Providing the input theorem, Sledgehammer first finds relevant premises using machine learning and a traditional filtering algorithm \cite{PAAR-2010:Three_Years_Experience_with}. Then it translates the input theorem together with a heuristic selection of hundreds of premises into goals that external ATPs can recognize. 
Then, ATPs solve these problems and return the results to Sledgehammer. Sledgehammer thus gets the useful premises and finally reconstructs proofs for the Isabelle trusted kernel to check. 
Every ITP has a trusted kernel, a minimal and rigorously verified core that checks manually constructed proofs following mathematical logic rules \cite{nipkow2002isabelle,pierce2010softwarefoundation}.
Since the kernel checks these proofs, it is unnecessary for the user to worry about the correctness of the generated proof.

Sledgehammer makes writing proofs convenient. Isabelle users only need to write the assumptions and the final goal that they want to prove. Then users can call Sledgehammer directly to finish the proof. Sledgehammer is good at generating short proofs \cite{PAAR-2010:Three_Years_Experience_with}. Thus, if users also provide a proof sketch with some subgoals introducing how the proof might be done, Sledgehammer will find proofs for each subgoal and construct the whole proof more easily. It is worth mentioning that Sledgehammer does not do induction very well. LLMs might help with theorems that require induction proofs.

\subsection{Explanation of Isabelle Code}
Listing \ref{listing:1} shows two Isabelle proofs of the same theorem, which states that reversing a list twice does not change the list. In the apply-style proof (right), we first apply induction on the list \textit{xs}, then we use a tactic \textit{auto} in Isabelle to finish the proof. The keyword \textit{done} marks the end of the proof. This proof style is concise, but to understand what these tactics are doing, we need to check the proof state of each step in Isabelle. In the Isar proof (left), we first apply induction on list \textit{xs} that clearly splits the goals into two cases, i.e., case \textit{Nil} and case \textit{Cons a xs}. Here \textit{Nil} means the empty list, and \textit{Cons} means appending an element \textit{a} to the front of the original list \textit{xs}. 

Note that these two styles of proofs can be combined, with Isar sketching the final proof while apply-style tactics discharge the remaining subgoals to fill in the sketch. For example, for the proof in Listing~\ref{listing:1}, a user could write a proof sketch in Isar by writing \textit{sorry} instead of \textit{by simp}---telling Isabelle to assume the subgoal is correct. Then, after using Isabelle to check that the proof sketch is sufficient, users can fill in the details to finish the proof.

\subsection{A Running Example of HybridProver}
\label{append:running example}

\begin{figure}[!htbp]
  \centering
  \includegraphics[width=\linewidth]{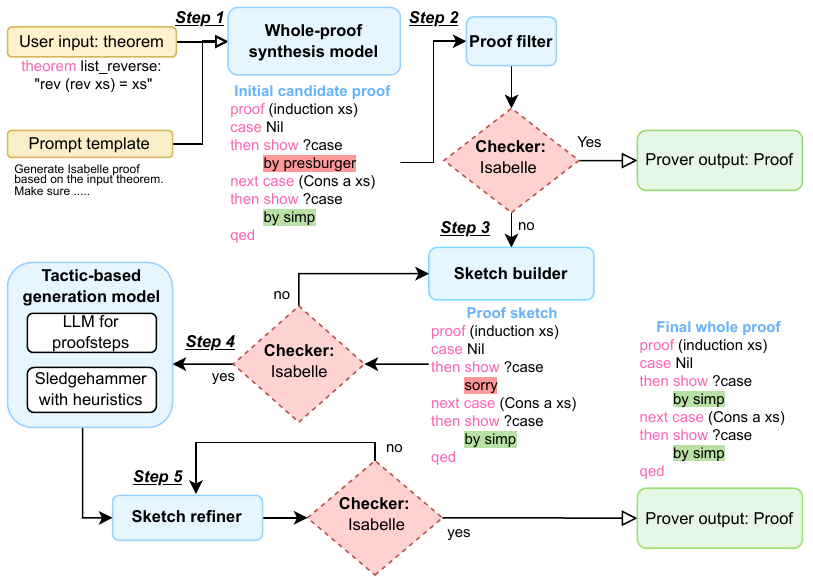}
  \caption{A running example of HybridProver}
  \label{fig:overview1}
\end{figure}
As a running example in Figure~\ref{fig:overview1}, we use the Isabelle theorem \textit{list\_reverse}, which states that reversing a list twice yields the original list:
\texttt{rev (rev xs) = xs}. 
The initial candidate proof generated by the whole-proof synthesis model follows an Isar-style induction structure. It starts with
\texttt{proof (induction xs)}, which splits the theorem into the base case \texttt{Nil} and the inductive case \texttt{Cons a xs}. 
In the candidate proof, the base case is discharged by \texttt{by presburger}, while the inductive case is discharged by \texttt{by simp}.

However, this candidate proof does not pass the Isabelle checker, because \texttt{by presburger} is not a suitable tactic for proving the \texttt{Nil} case of this list theorem. 
Although the overall proof structure is reasonable---it correctly uses induction and introduces the two standard list cases---one of its local proof steps is incorrect. 
This example illustrates why directly discarding the whole candidate proof would waste useful structure.

HybridProver therefore constructs a proof sketch from this failed candidate. 
In the sketch shown in Figure~\ref{fig:overview1}, the concrete tactics \texttt{by presburger} are replaced with \texttt{sorry}, while the correct subgoal proof and surrounding Isar structure is preserved:
\texttt{proof (induction xs)}, \texttt{case Nil}, \texttt{next case (Cons a xs)}, and \texttt{qed} remain unchanged. 
The resulting sketch keeps the high-level induction argument but leaves the local proof obligations unresolved.

During refinement, the tactic-based generation model and Sledgehammer search for concrete tactics to replace the \texttt{sorry} placeholders. 
For this example, the first subgoal can be solved by \texttt{by simp}. 
After substitution, the final proof replaces each \texttt{sorry} with \texttt{by simp}, producing a complete Isabelle proof that passes verification:
the base case \texttt{Nil} is proved by simplification, and the inductive case \texttt{Cons a xs} is also discharged by simplification.

This example shows that the initial whole-proof candidate already contains a useful high-level structure, even though one local tactic is wrong. 
By preserving the Isar structure and refining only the unresolved proof steps, HybridProver converts the failed candidate into a verified proof.


\subsection{Additional Related Work}

\paragraph{Interactive theorem proving.}
Theorem proving is widely used for ensuring the safety and security of critical software and systems. Most work in interactive theorem proving involves some manual proof construction by human experts, a process that is labor-intensive and time-consuming. 
For example, the seL4 team \cite{klein2009sel4} takes 30 person-years to verify a general-purpose microkernel with 8700 lines of C code in Isabelle.
The CertiKOS team \cite{gu2016certikos} verifies a concurrent OS through deep abstraction layers in Rocq \cite{gu2015deep,gu2018certified}.
They take 3 person-years to prove 6500 lines of C and x86 assembly. 
CompCert \cite{leroy2016compcert} is the first formally verified C compiler to be exempt from miscompilation. Six person-years are taken in Rocq to construct proofs.
Additionally, most theorem-proving works in other operating systems take more than 1 person-year \cite{sigurbjarnarson2016push,shinde2020besfs,chajed2019verifying,zhao2017refinement,amani2016cogent,proverit_hu,verbeek2015formal,11106232}.
Automated theorem provers (ATPs) such as Z3 and CVC4 offer full automation~\cite{barrett2011cvc4,de2008z3,1218615.1218620,weidenbach1996spass}, but are typically limited to first-order logic and suffer from search-space explosion. 
This motivates the use of LLMs to improve the usability of ITPs without sacrificing their expressiveness and scalability.

\paragraph{Autoformalization.} In this work, we focus on theorem proving where a formal specification already exists. Another interesting area is autoformalization, where researchers translate natural language texts into formal specifications and proofs  \cite{azerbayev2023proofnet,wang2018first,cosler2023nl2spec,hahn2022formal,wu2022autoformalization,chen2023nl2tl}. 
The amount of natural language texts and mathematical proofs is far greater than formal texts. Therefore, autoformalization creates a lot of high-quality data used for model training, which solves the data scarcity problem. Other works also try to generate specifications based on code \cite{wen2024enchanting,cosler2023nl2spec,si2018learning,si2020code2inv}. 

\paragraph{Premise Selection.} Premises refer to the auxiliary lemmas, as well as datatype and function definitions, that are referenced to establish the target theorem. Existing studies on premise selection mainly follow two technical routes. One line of work builds an external premise retrieval library and retrieves relevant premises according to semantic similarity or symbolic matching with the input theorem~\cite{wang2024lego,yang2023leandojo}. The other incorporates premise usage into model training, enabling the model to memorize common premises and their contexts and to invoke them automatically when proving similar theorems ~\cite{first2023baldur}. We adopt the latter approach by integrating premise selection into the training objective of the whole-proof synthesis model. As a result, the model is encouraged to identify and insert necessary premises during proof generation, thereby providing richer contextual support for the subsequent strategy generation stage and improving the accuracy of subgoal solving.

\subsection{Additional Setup}
\label{append:sft and rl}
\paragraph{Dataset.}
PISA includes the STD and AFP with version 2021-10-22. The former is more foundational, with around 70K theorems. The latter is a proof library contributed by most Isabelle researchers, with around 170K theorems. It diverges from mathematical problems to formal verification of software and hardware.

MiniF2F \cite{zheng2022minif2f} is
used in many research studies \cite{yang2023leandojo,jiang2023draft,zhao2024subgoalxl}. It is an open-source collection containing 488 high school math competition problems. These problems are divided into two groups: 244 for validation and 244 for final testing. All problems have been translated into 4 proof assistant languages: Lean, HOL light, Metamath, and Isabelle.

\paragraph{Isabelle Interactive environment.}
We interact with Isabelle using Isabelle-client \cite{10.1007/978-3-031-16681-5_24}, which provides detailed developer documentation and Isabelle interface descriptions, making it easy to use and maintain. It is a Python client library that strictly follows the Isabelle system manual \cite{wenzel2021isabelle}. To improve system throughput and efficiency, we employ a concurrent architecture based on a single server, a single client, and multiple sessions. Our experiments show that using multiple Isabelle server processes leads to resource contention. By contrast, multiple independent sessions can load different Isabelle theory libraries separately, thereby supporting the verification of different theorems in the evaluation dataset. In addition, we found that a multi-client single-session setting is error-prone and incurs high resource consumption, and therefore it is not adopted in our system.

\textit{watchdog\_timeout} is set to 30s for evaluation. Evaluation in Isabelle is time-consuming even for a small sampling rate and a small dataset. It costs 300 CPU hours to evaluate 244 theorems in miniF2F with a 32 sampling rate for one LLM. Thus, we do ablation studies with a relatively low sampling rate and small timeout. To set up Sledgehammer, we follow the default Isabelle2021 configuration \cite{jiang2021lisa}, where the Sledgehammer timeout limit is 30s, with the on-machine five default ATPs (E, SPASS, Vampire, Z3, and CVC4). 
We also complement Sledgehammer with 11 heuristics for proofs: \textit{['by auto', 'by simp', 'by blast', 'by fastforce', 'by force', 'by eval', 'by presburger', 'by sos', 'by arith', 'by linarith', 'by (auto simp: field\_simps)']}. 

\paragraph{Training Configurations.}

For fine-tuning, we use LoRA \cite{hu2022lora} to train around 1\% of all parameters with a \textit{lora\_rank} 32 and all target layers open for adjustment.  
In the first 10\% of steps, the learning rate warms up linearly from 0 to the maximum value $e^{-5}$. Then it decays to 0 following a cosine schedule. The context length for the training data is 2048 tokens, and the total batch size is 64.

We implement reinforcement learning using the GRPO trainer from the TRL framework~\cite{vonwerra2020trl}. For each input theorem, we sample 16 candidate proofs in parallel. The reward function combines two components: (1) a verification reward obtained from Isabelle, and (2) a penalty for undesirable patterns. Concretely, proofs that pass Isabelle checking receive a reward of 1.0, proofs that are syntactically valid but fail verification receive 0.2, and invalid outputs receive 0.0. Additionally, outputs containing tokens such as \texttt{sorry} or malformed proof patterns are assigned a negative penalty (-0.2), discouraging trivial or incomplete proofs.
We train the model for 3 epochs with a learning rate of $2\times10^{-5}$, using a per-device batch size of 2.

\subsection{Prompt template and few-shot examples}

\begin{mylisting}[label=listing:prompt]{Prompt template}
Generate proof code in Isabelle/HOL based on the input Isabelle/HOL theorem. Make sure that the generated proof can be verified by Isabelle/HOL. Do not generate extra natural language descriptions.
**Input theorem:**
\end{mylisting}

\begin{mylisting}[label=listing:few-shot, listing options={language=Isabelle}]{Few-shot examples}
|\ttfamily\color{skyblue}Input theorem 1:| 
theorem app_Nil: "[] @ xs = (xs :: 'a list)"
|\ttfamily\color{skyblue}Isabelle proof 1:| 
{|\ttfamily\color{black}Isabelle proof|: by simp}

|\ttfamily\color{skyblue}Input theorem 2:| 
theorem list_reverse: "rev (rev xs) = xs"
|\ttfamily\color{skyblue}Isabelle proof 2:| 
{|\ttfamily\color{black}Isabelle proof|: apply (induct xs) apply simp apply simp done}

|\ttfamily\color{skyblue}Input theorem 3:| 
theorem append_assoc: "(xs @ ys) @ zs = xs @ (ys @ zs)"
|\ttfamily\color{skyblue}Isabelle proof 3:| 
{|\ttfamily\color{black}Isabelle proof|: proof (induct xs) case Nil then show ?case by simp next case (Cons x xs) then show ?case by simp qed}


\end{mylisting}

\subsection{Additional ablation results}
\paragraph{RQ4: Is the proof pattern similarity equal to the proof quality?}

During the fine-tuning stage, we try to feed the model enough theorem-proof pairs. The LLM learns to generate proof patterns similar to the training data. But is the proof pattern similarity equal to the quality? Thus, we not only test the success rate of the miniF2F dataset, but also record all responses from the sampling during the whole evaluation. We make a comparison between them and find that generally, the higher the total correct response is, the higher the success rate for all theorems is. It demonstrates that proof pattern similarity is strongly relevant to the proof quality, reflecting the improvement from training.

\begin{tcolorbox}[
  colback=gray!10,
  colframe=gray!80,
  boxrule=1.5pt,
  arc=4pt,
  left=6pt,
  right=6pt,
  top=6pt,
  bottom=6pt
]
\textbf{RA4:} The more similar an LLM-generated proof pattern is to the training data, the higher the quality of the proof.
\end{tcolorbox}


\end{document}